\documentclass[3p,times]{elsarticle}

\usepackage{ecrc}

\volume{00}

\firstpage{1}

\journalname{Nuclear Physics A}

\runauth{Kari J. Eskola}

\jid{nupha}

\usepackage{amssymb}

\usepackage[figuresright]{rotating}

\begin{document}

\begin{frontmatter}



\dochead{}

\title{Global analysis of nuclear PDFs -- latest developments}


\author{Kari J. Eskola}

\address{Department of Physics, P.O. Box 35, FI-40014 University of Jyv\"askyl\"a, Finland\\
Helsinki Institute of Physics, P.O. Box 64, FI-00014 University of Helsinki, Finland}

\begin{abstract}
In this review talk I discuss the latest developments in the DGLAP-based global analysis of nuclear parton distribution functions (nPDFs), focusing on the recent EPS09, nCTEQ and DSSZ global fits.  I also briefly review the recent analysis for assigning a spatial dependence to the globally analysed nPDFs, resulting in the new sets EPS09s and EKS98s ("s" for spatial). With these, one can now compute nuclear hard-process cross sections and estimate their nPDF-originating uncertainties in different centrality classes for the first time consistently with the global nPDF fits. 
\end{abstract}


\end{frontmatter}



\section{Introduction: Global DGLAP analyses of nPDFs}
\label{sec:intro}

Inclusive cross sections of hard processes involving a large interaction scale $Q\gg\Lambda_{\rm QCD}$ 
in a high-energy hadronic or nuclear collision of particles $A$ and $B$  can be computed using the QCD collinear factorization theorem, \begin{equation}
\mathrm{d} \sigma^{AB \rightarrow k + X} = \sum\limits_{i,j,X'} f_{i}^A(Q^2) \otimes f_{j}^B(Q^2) \otimes  \mathrm{d}\hat{\sigma}^{ij\rightarrow k + X'} + {\cal O}(1/Q^2), 
\label{eq:sigmaAB1}
\end{equation}
where $\mathrm{d}\hat{\sigma}$ are the perturbatively calculable partonic pieces, and $f_i^{A}(f_j^{B})$ is the universal, process-independent, parton distribution function (PDF) for a parton flavor $i(j)$ in $A(B)$. The PDFs are of nonperturbative origin, but their evolution in the scale $Q^2$ can be obtained from the DGLAP equations \cite{Dokshitzer:1977sg,Gribov:1972ri,Gribov:1972rt,Altarelli:1977zs} derived from perturbative QCD. 

A precise knowledge of the PDFs is vital for interpreting any hard-process results in p+p, p+$A$, d+$A$, and $A$+$A$ collisions
at the present BNL-RHIC and CERN-LHC colliders. Consequently, global analyses which exploit a multitude of experimental hard-process data and the DGLAP evolution, have been developed to determine the nonperturbative input in the PDFs. Traditionally, in these analyses the PDFs are parametrized at some perturbative initial scale, $Q=Q_0\gg \Lambda_{\rm QCD}$, where the power corrections, ${\cal O}(1/Q^2)$ in Eq.~(\ref{eq:sigmaAB1}), can be expected to be small. Performing the DGLAP evolution to higher scales, and computing the cross sections of the measured observables (with each parameter candidate-set at the time), one then iteratively determines the parameter set giving the best fit with the measured data. A statistical error analysis is performed after this, to determine how the experimental statistical and systematic errors translate into uncertainties of the PDFs. With the resulting error sets and best fit for the PDFs, one may then compute how the uncertainties in the PDFs propagate into the hard-process cross sections. Excellent fits for the free-proton PDFs have been obtained in this way, resulting in  the sets like CT10 \cite{Lai:2010vv} and MSTW \cite{Martin:2009iq}. Also neural network techniques have been successfully developed, resulting in  e.g. the set NNPDF2.0 \cite{Ball:2010de}. 

Challenges for the global PDF analyses are posed, e.g., by the treatment of experimental errors and weighting of those data sets which contain a small number of datapoints but which still offer very valuable constraints. Due to the large dimension of the fit-parameter space, ${\cal O}(15-30)$, and due to the complexity of the NLO (and NNLO) cross sections,  very fast solvers for the DGLAP evolution and cross sections are also needed. 
For the global analysis of nuclear PDFs, the main further challenge is essentially the fact that we have fewer types of data and
less data points covering a much more limited $x,Q^2$ region at our disposal, i.e. less constraints available than in the free proton case. 
In addition, one needs to account for the $A$ dependence, and also the spatial (impact parameter) dependence of the nPDFs. All the global analyses so far, listed in Table~1, have been performed for the spatially averaged nPDFs probed in minimum-bias nuclear collisions with no cuts on the collision centrality (impact parameter). An obvious weakness of these nPDFs then is that it has not been possible to consistently compute nuclear hard-process cross-sections in different centrality classes.
\begin{table}[tbh]
\begin{center}
\begin{tabular}{llllll}
\hline
year &set &Ref. & order & data types used & error analysis \\
\hline
1998 & EKS98 & \cite{Eskola:1998iy,Eskola:1998df} & LO & $l$+$A$ DIS, p+$A$ DY & no \\
2001 & HKM   & \cite{Hirai:2001np}                & LO & $l$+$A$ DIS 						& \underline{yes}\\
2004 & HKN04 & \cite{Hirai:2004wq}								 & LO & $l$+$A$ DIS, p+$A$ DY & yes\\ 
2004 & nDS   & \cite{deFlorian:2003qf} 					 & \underline{NLO}& $l$+$A$ DIS, $p$+$A$ DY	& no \\
2007 & EKPS  & \cite{Eskola:2007my}							 & LO & $l$+$A$ DIS, p+$A$ DY & yes\\
2007 & HKN07 & \cite{Hirai:2007sx} 							 & NLO& $l$+$A$ DIS, p+$A$ DY & \underline{yes}\\
2008 & EPS08 & \cite{Eskola:2008ca}							 & LO & $l$+$A$ DIS, p+$A$ DY, \underline{$h^{\pm},\pi^0,\pi^{\pm}$ in d+Au} &no\\
2009 & \textbf{EPS09} & \cite{Eskola:2009uj} 							 & NLO& $l$+$A$ DIS, p+$A$ DY, $\pi^0$ in d+Au & yes, $\rightarrow$ \underline{error sets}\\
2009 & nCTEQ & \cite{Schienbein:2009kk,Stavreva:2010mw} 
                                                   & NLO& $l$+$A$ DIS, p+$A$ DY 					& yes\\
2010 & \textbf{nCTEQ} & \cite{Kovarik:2010uv}							 & NLO& $l$+$A$ and \underline{$\nu$+$A$} DIS, p+$A$ DY & yes\\
2012 & \textbf{DSSZ}  & \cite{deFlorian:2011fp} 					 & NLO& $l$+$A$ and $\nu$+$A$ DIS, p+$A$ DY, &\\ 
&&&&$\pi^0,\pi^{\pm}$ in d+Au, \underline{computed with nFFs} & yes, $\rightarrow$ error sets\\
\hline
\label{tab:nPDFs}
\end{tabular}
\end{center}
\vspace{-0.5cm}
\caption{The developments in the global DGLAP analysis of nPDFs since 1998. The new elements are indicated with underlining. DIS = deep inelastic scattering; DY = Drell-Yan dilepton production; nFFS = nuclear fragmentation functions; bold face = reviewed this talk.}
\end{table}

Table 1 summarizes the key developments in the global analysis of the nPDFs. Next, I will briefly review the most recent ones of these. 
The impact-parameter dependent nPDFs of Ref.~\cite{Helenius:2012wd} (see also Helenius  \cite{Helenius:2012wk} in these Proc.) will be discussed after this. I will not address the possible power corrections in the cross sections \cite{Qiu:2003vd}, or nonlinearities in the scale evolution, gluon saturation phenomena or other evolution equations than DGLAP here. It is also good to keep in mind that the global DGLAP analyses of the nPDFs do not address the interesting question of the origin of the nuclear effects. For such studies, see e.g. the modeling in Refs. \cite{Frankfurt:2003zd,Tywoniuk:2007xy,Guzey:2009jr} and the reviews in Refs. \cite{Armesto:2006ph,Frankfurt:2011cs}.

\section{EPS09}
\label{sec:EPS09}

\begin{figure}[thbp]
\centering
\vspace{-.2cm}\hspace{-0.5cm}
\includegraphics[width=10cm]{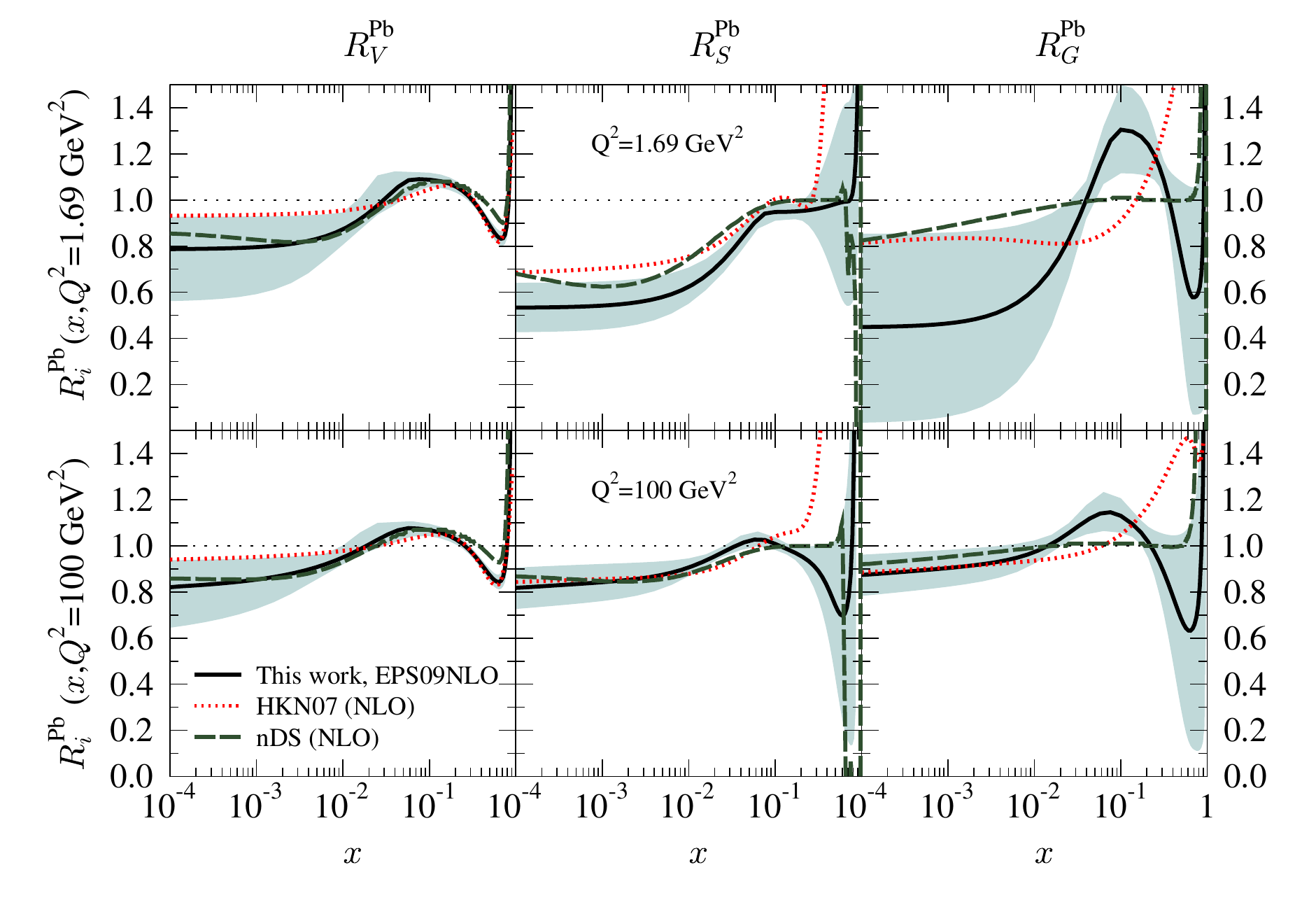} \hspace{-0.cm}
\includegraphics[width=6.5cm]{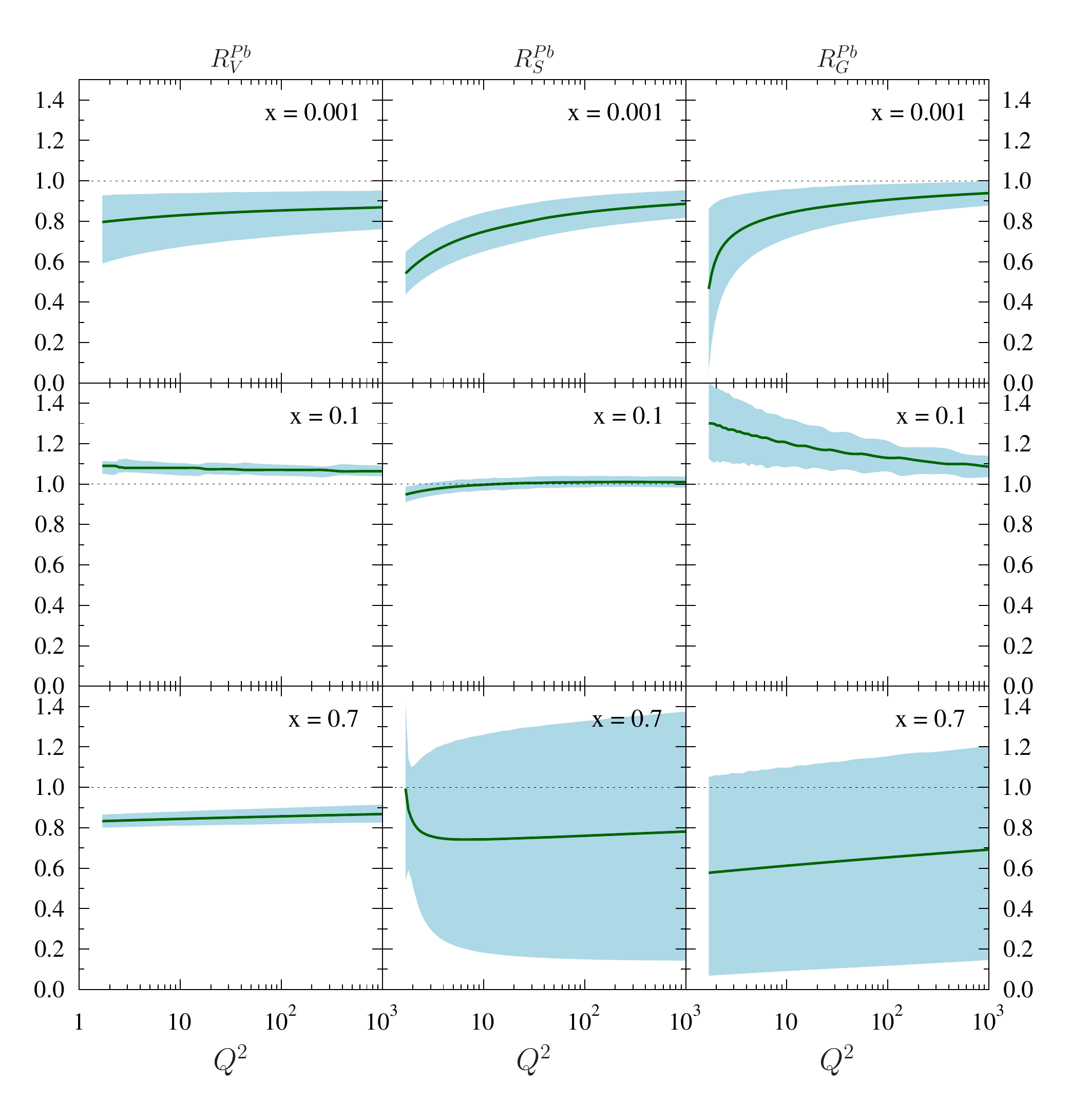}\hspace{-0.5cm}
\vspace{-0.7cm}
\caption{(Color on-line) Left: The nuclear effects to bound-proton NLO PDFs in a lead nucleus as a function of $x$, at two different scales, according to the EPS09 global analysis. Comparison with the earlier NLO nPDFs is also shown. The error bands are computed from the 15+15 error sets obtained through a "90\% confidence criterion". From  \cite{Eskola:2009uj}.
Right: $Q^2$ evolution of the EPS09 nuclear effects at fixed values of $x$. From \cite{Eskola:2010jh}.
}
\label{fig:EPS09}
\end{figure}
\begin{figure}[t!]
\centering
\vspace{-.2cm}
\includegraphics[width=9cm]{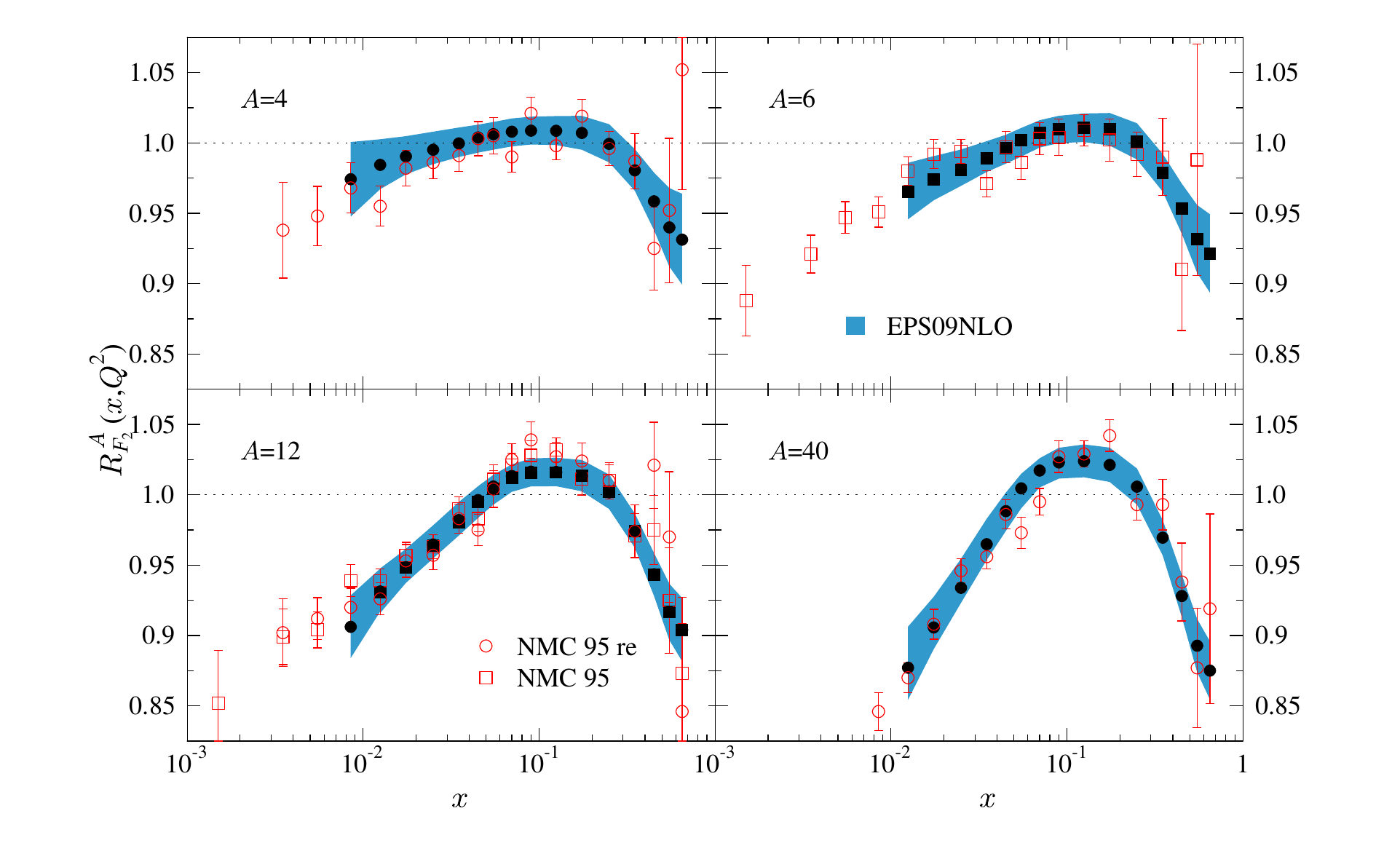} \hspace{0cm}
\includegraphics[width=6cm]{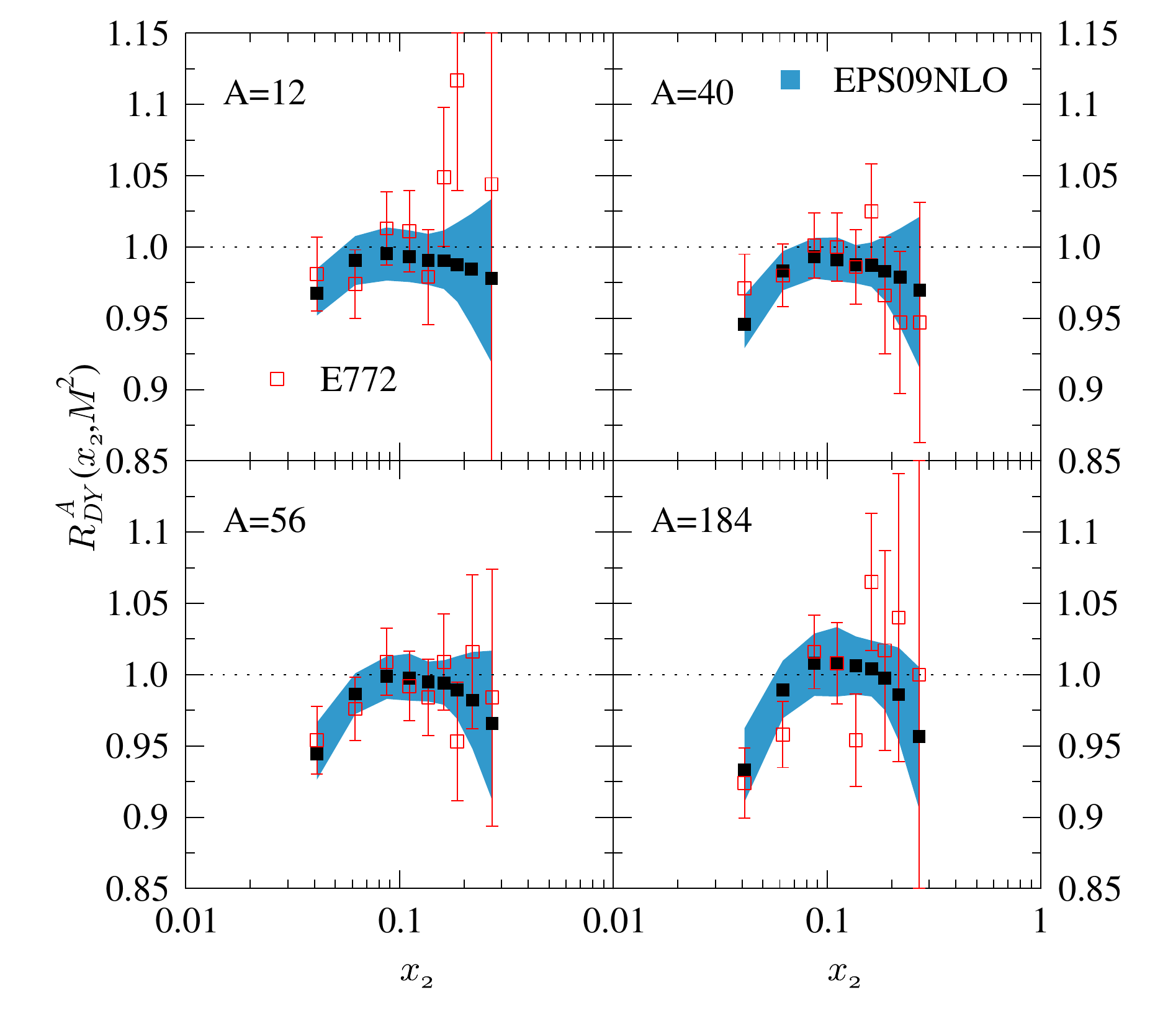}\hspace{0cm}

\vspace{-0.3cm}
\includegraphics[width=7cm]{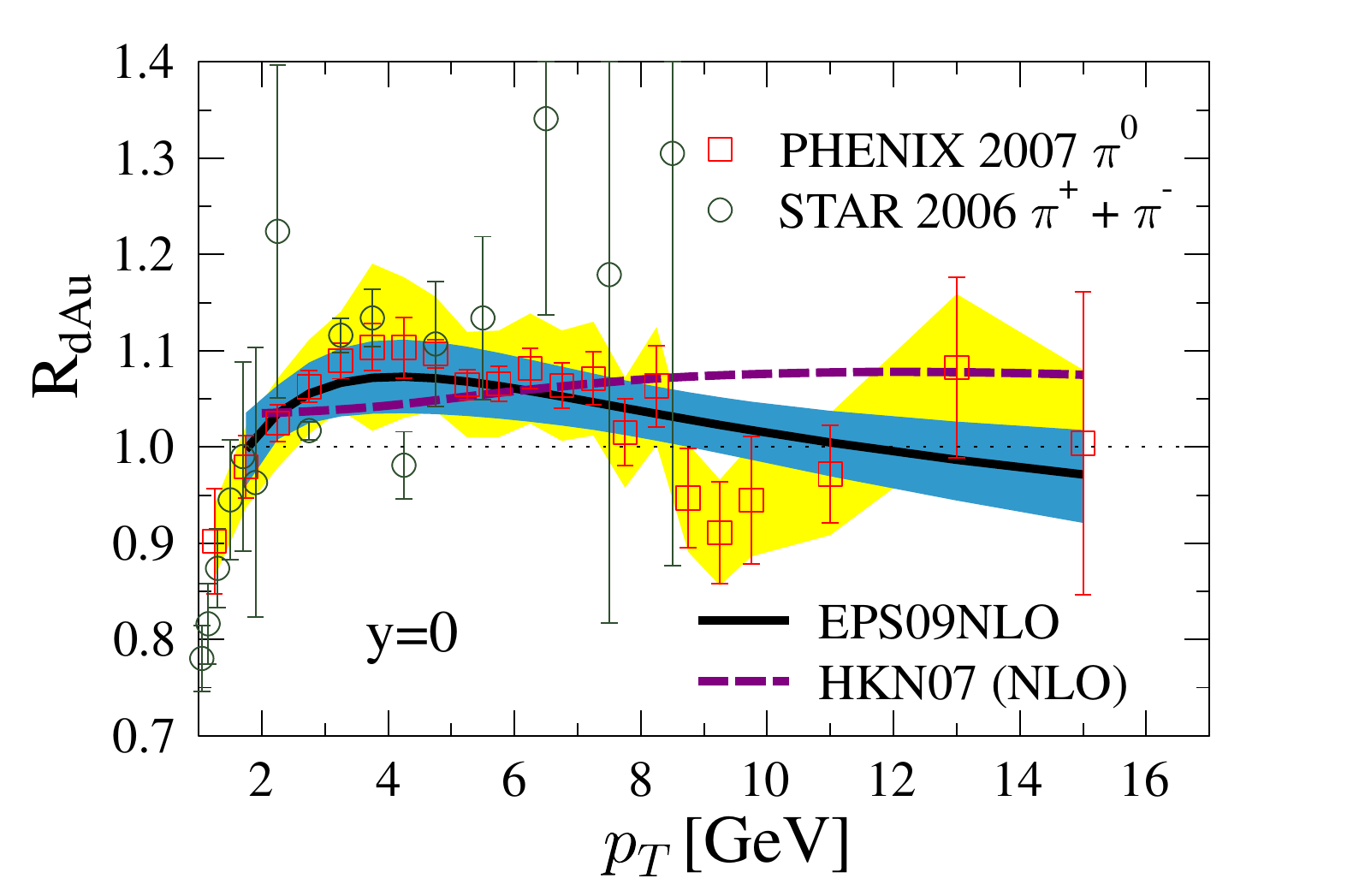}
\vspace{-0.3cm}
\caption{(Color on-line) The EPS09 fit to some of the $l$+$A$ DIS $\frac{1}{A}F_2^A(x,Q^2)/\frac{1}{2}F_2^{\rm d}(x,Q^2)$ data (upper left), p+$A$ DY  $(\frac{1}{A}d\sigma^{{\rm p}A}/dx_2dQ^2)/(\frac{1}{2}d\sigma^{\rm pd}/dx_2dQ^2)$ data (upper right) sets used, as well as to the minimum-bias PHENIX nuclear modification factor data for $\pi^0$ production in d+Au collisions, $\frac{1}{2A}(d\sigma^{\rm dAu}/dp_Tdy)/(d\sigma^{\rm pp}/dp_Tdy)$, in mid-rapidity at RHIC \cite{Adler:2006wg} (lower panel). The $\pi^0$ production is computed with the INCNLO code \cite{Aversa:1988vb} and NLO KKP vacuum fragmentation functions (FFs) \cite{Kniehl:2000fe}.
From \cite{Eskola:2009uj}.}
\label{fig:DY_DIS_pi0}
\vspace{-0.3cm}
\end{figure}

Following the EKS98 framework \cite{Eskola:1998iy,Eskola:1998df}, in the EPS09 analysis \cite{Eskola:2009uj} the (spatially independent) bound-proton nPDFs are defined in terms of the nuclear modifications, $R_i^{A}(x,Q^2)$, and free-proton PDFs, $f_i^p(x,Q^2)$: 
\begin{equation}
f_i^{p/A}(x,Q^2) \equiv R_i^{A}(x,Q^2) f_i^p(x,Q^2).
\end{equation}
The bound-neutron PDFs are obtained by assuming isospin symmetry. As the free proton baseline, EPS09 uses the CTEQ6.1M free proton NLO PDFs \cite{Pumplin:2002vw}. Accordingly, the initial scale in EPS09 is set to $Q_0=1.3$~GeV, and the MSbar regularization scheme and zero-mass variable flavour-number heavy-quark scheme (i.e. heavy quarks are taken massless and generated only radiatively through the DGLAP evolution above the mass thresholds) are adopted. Uncertainties in the free-proton PDFs are not considered. As the constraints, EPS09 exploits 25 data sets from $l$+$A$ DIS [E139 and NMC experiments],  6 sets from DY in p+$A$ [E772 and E866], and one set from inclusive minimum-bias $\pi^0$ production in d+Au collisions at RHIC [PHENIX], altogether 929 datapoints (for the Refs., see \cite{Eskola:2009uj}). The quality of the EPS09 best fit is excellent: $\chi^2/{\rm datapoints}=0.79$. The statistical error analysis is done through the Hessian method, by first diagonalizing the Hessian matrix and finding the parameter eigendirections as linear combinations of the 15 fit parameters. The 15 + 15 error sets, pairwise assigned to each of the uncorrelated eigendirections, are determined with a "90\% confidence criterion" (see the details in \cite{Eskola:2009uj}) which in the EPS09 case is equivalent to allowing the $\chi^2$ to vary by $\Delta\chi^2=50$, i.e. ca. 5\%. The resulting error band and comparison with the earlier NLO nPDFs is shown in Fig.~\ref{fig:EPS09} for the lead nucleus. The good simultaneous fit to the DIS, DY and $\pi^0$ data is demonstrated in Fig.~\ref{fig:DY_DIS_pi0}.
As seen in Fig.~\ref{fig:EPS09}, large uncertainties remain especially in the small-$x$ and large-$x$ gluons. The DGLAP evolution diminishes the small-$x$ gluon shadowing and also its error band, as seen in the lower panel. The decisive role of the $\pi^0$ data \cite{Adler:2006wg} in fixing the mid-$x$ EMC-effect of gluons is demonstrated in the lower panel of  Fig.~\ref{fig:DY_DIS_pi0}. 
As a summary of the EPS09 analysis, we can conclude that the excellent simultaneous fairly tensionless fit to the different types of nuclear hard-process data suggests that factorization works well in the energy range studied, and that the extracted nPDFs seem universal in the region $x\gtrsim0.005$, $Q\geq 1.3$~GeV. Also the procedure for propagating the experimental errors into the nPDF uncertainties and error sets is seen to work well.

\section{nCTEQ}
\label{sec:nCTEQ}
The latest nCTEQ global analysis for nPDFs, presented in Ref. \cite{Kovarik:2010uv}, is built on a similar set-up as the 
CTEQ6M free proton PDFs. Instead of the nuclear modifications to the PDFs, nCTEQ parametrizes directly the initial bound-proton PDFs, $f_i^{p/A}(x,Q_0^2)$, at a the scale $Q_0=1.3$~GeV. Heavy-quark effects are included here in a more sophisticated way, using the general-mass variable-flavour number scheme (GM-VFNS). Altogether 31 charged-lepton+$A$ DIS data sets [E139, NMC, EMC, E665, BCDMS, HERMES] and 6 DY p+$A$ data sets [E772, E886], consisting of altogether 708 datapoints were used as constraints to the nPDFs. As a new element relative to EPS09, nCTEQ includes also 8 $\nu(\bar\nu)$+$A$ DIS data sets [CHORUS, NuTeV, CCFR] with as many as 3134 datapoints into the global analysis. The RHIC data have not, however, been used here at all. For all the data Refs., see \cite{Kovarik:2010uv}.

The interesting main observation from the nCTEQ analysis is essentially shown by Table~2: with increasing weight $w$ to the neutrino-DIS data, the $l^{\pm}$-DIS\&DY fit and global fit deteriorate quite significantly, which indicates a tension between the $l^{\pm}$-DIS\&DY data and the $\nu$-DIS data. We can also see that the fit to the $\nu$-DIS data alone is of a poorer quality than that to the $l^{\pm}$-DIS\&DY data alone. Based on the detailed statistical analysis, the nCTEQ results in \cite{Kovarik:2010uv} would seem to suggest that the nPDFs are not universal but process-dependent. As pointed out in \cite{Kovarik:2010uv,Kovarik:2012pm}, this conclusion depends, however, on the analysis details, in particular on the extent to which the systematic errors of the $\nu$-DIS data are regarded correlated. Namely, as pointed out in \cite{Paukkunen:2010hb}, where these errors were taken uncorrelated, no such a strong conclusion was found. In addition, in Ref.~\cite{Paukkunen:2010hb} it was demonstrated that while a relatively good agreement with the CHORUS and CDHSW data was found, 
the NuTeV data, and especially the fluctuations in the normalization of the data from one neutrino energy to another, causes the worsening of the fits. This tension problem with the NuTeV data alone (which form  ca. 2/3 of all the $\nu$-DIS data) was also acknowledged in \cite{Kovarik:2012pm}. At this point, my conclusion is that factorization seems to work for the nPDFs, even with the $\nu$-DIS data included.
To more stringently test the universality of the nPDFs, obviously more precise $\nu$-DIS data would be needed in the future. 
\begin{table}[tbh]
\begin{center}
\begin{tabular}{llll}
\hline
$w$ & $l^{\pm}$-DIS \& DY   & $\nu^{\pm}$-DIS  				& total $\chi^2/{\rm datapt}$\\
\hline
0 		& 0.9	 									& -												& 0.90			\\
1/2 	& 0.96 								 	& 1.40										& 1.32 			\\ 
1 		& 1.04   								& 1.36 					 					& 1.30 			\\
$\infty$ & -				 					& 1.33  									& 1.33			\\
\hline
\end{tabular}
\end{center}
\vspace{-0.3cm}
\caption{The worsening of the global fit as a function of the weight $w$ to the neutrino-DIS data in the nCTEQ analysis. 
From Ref. \cite{Kovarik:2010uv}.} \label{tab:nCTEQ_nPDFs}
\vspace{-0.3cm}
\end{table}

\section{DSSZ}
\label{sec:DSSZ}
The latest global analysis for the nPDFs is presented by DSSZ in Ref.~\cite{deFlorian:2011fp}.  
Here the baseline free-proton PDF set is MSTW \cite{Martin:2009iq}, and  the nuclear modifications $R_i^A(x,Q_0^2)$ of the bound protons are parametrized at an initial scale $Q_0=1$~GeV. The MSbar regularization and GM-VFNS heavy-quark scheme are used. As constraints for the nPDFs, DSSZ exploits 27 $l^{\pm}$-DIS data sets [NMC, E139, EMC], 6 p+$A$ DY data sets [E772, E866], together with 6 $\nu$-DIS data sets [NuTeV, CDHSW, CHORUS] and 3 pion data sets in d+Au at RHIC [PHENIX,STAR], summing up to 1579 datapoints altogether (for the Refs., see \cite{deFlorian:2011fp}). The DSSZ analysis thus covers the most extensive selection of the nuclear hard-process data types so far. The key features, and simultaneously also the key differences relative to EPS09, are that no weights are introduced for the data sets, and that in computing the inclusive pion production in d+Au collisions at RHIC, DSSZ uses the \textit{nuclear} fragmentation functions (nFFs) determined in \cite{Sassot:2009sh}. The effects of uncertainties in the free-proton MSTW PDFs are not examined in nuclear ratios but DSSZ cleverly includes these as additional errors to the measured absolute $\nu$-DIS structure functions. The DSSZ analysis arrives at a very good overall fit, $\chi^2/{\rm datapoint}=0.98$. The propagation of the experimental uncertainties into the nPDFs is quantified via the Hessian analysis, which yields 25+25 error sets corresponding to a choice $\Delta \chi^2=30$ (i.e. allowing for a 2\% increase of $\chi^2$). The conclusion from the DSSZ analysis is that good simultaneous and tensionless fits to the data used are found, and, in particular, there seems to be no conflict between the nuclear modifications of the $l^{\pm}$-DIS and $\nu$-DIS data. Also based on DSSZ, the measured nuclear effects in hard processes can be explained by universal nPDFs.   
For a detailed account of the DSSZ nuclear modifications and their error bands, see Figs.~13-17 in Ref.~\cite{deFlorian:2011fp}. A comparison of the DSSZ with the EPS09 and nDS nuclear effects at a scale $Q^2=10$~GeV$^2$ is presented in Fig.~14 of Ref.~\cite{deFlorian:2011fp}.
\begin{figure}[htbp]
\centering
\vspace{-0.3cm}
\includegraphics[width=13cm]{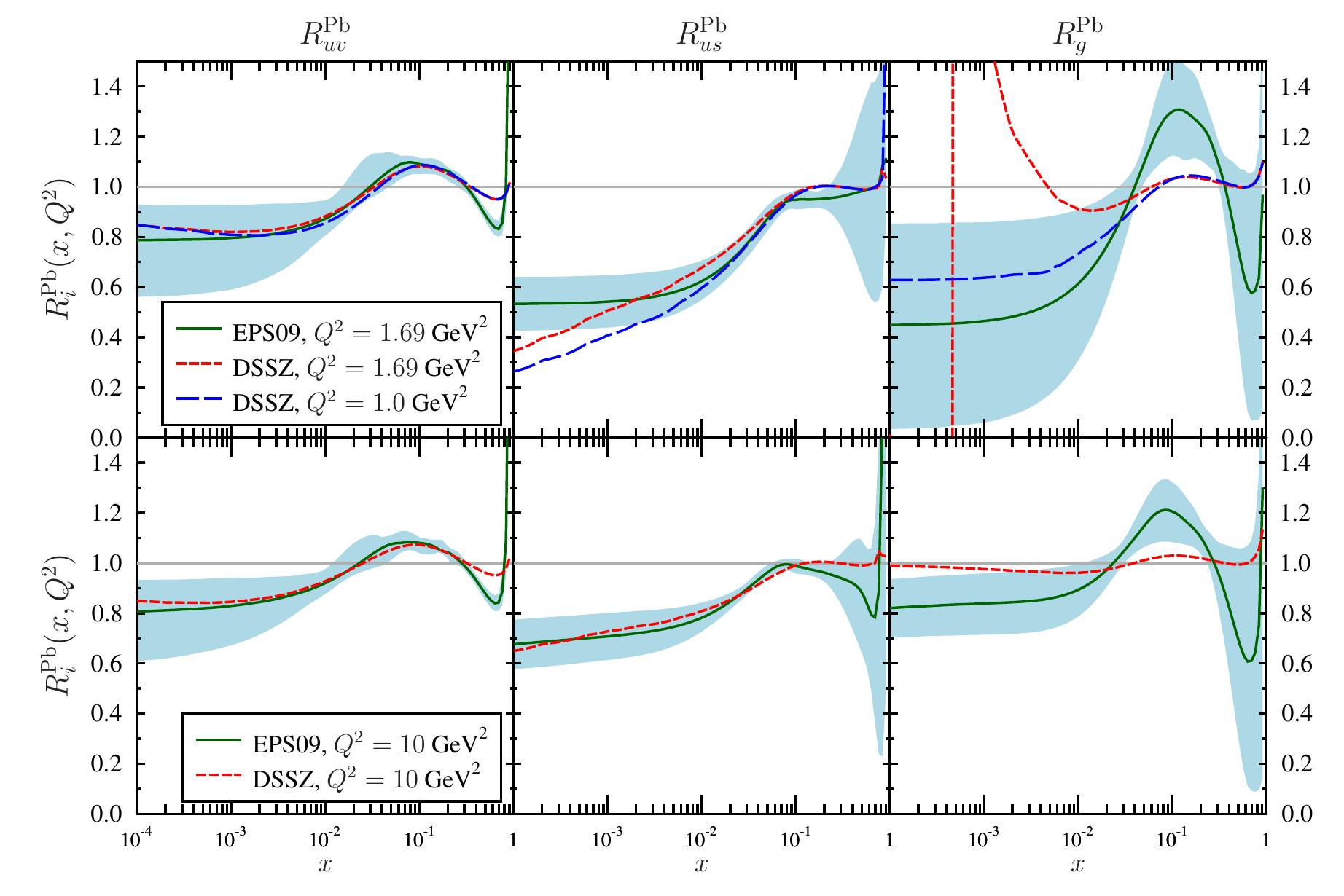}
\vspace{-0.4cm}
\caption{(Color on-line) Upper panels: The DSSZ and EPS09 nuclear effects to bound-proton PDFs (valence $u$ quarks, sea $u$ quarks, gluons) in a lead nucleus as a function of $x$ at the initial scales $Q_0^2=1.0$~GeV$^2$ for DSSZ (blue long dashed) and $1.69$~GeV$^2$ for EPS09 (green solid). The error bands are from EPS09. The DSSZ modifications at $Q^2=1.69$~GeV$^2$ are also shown (red short dashed); their complicated behaviour is caused by the negative gluon PDFs in MSTW and DSSZ at smallest $x$ and small $Q^2$. 
Lower panels: The same but for $Q^2=10$~GeV$^2$. Prepared by I. Helenius.}
\label{fig:DSSZvsEPSinx}
\end{figure}

It is interesting to compare the EPS09 and DSSZ results, see Fig.~\ref{fig:DSSZvsEPSinx} above, especially in the gluon shadowing and antishadowing regions, where the results from these two analyses differ the most. 
The initial nuclear modifications $R_g^A(x,Q_0^2)$ in the smallest-$x$ region which is not directly constrained by the existing nuclear hard-process data, are fairly similar in DSSZ (long dashed)  and in EPS09 (solid) but their different initial scales should be noticed. The negative smallest-$x$ gluon PDFs at low scales both in MSTW and in DSSZ (which as such are not problematic, the PDFs themselves evolve in a stable manner and the physical cross-sections remain of course positive) combined with the rapid DGLAP evolution of the small-$x$ gluons at $Q\sim 1$~GeV obscure the interpretation of the $R_g^A(x,Q^2)$ in DSSZ at the smallest scales above $Q_0$. As shown in the upper right panel on the right in Fig.~\ref{fig:EPS09} illustrating these scale evolution effects, with the non-negative gluons in EPS09 and CTEQ6M these interpretation (and control) problems do not occur.

As seen in Fig.~\ref{fig:DSSZvsEPSinx}, in the valence and sea quark sectors the DSSZ and EPS09 do not differ significantly. At the smallest $x$, the sea-quark differences reflect the lack of small-$x$, high-$Q^2$ nuclear DIS data. For the valence quarks, there is a difference in the mid-$x$ EMC-region, the origin of which should be clarified.

Figure~\ref{fig:DSSZvsEPSinx} also shows that the amounts of gluon antishadowing at $x\sim 0.1$ in DSSZ and EPS09 are quite different, and while the EPS09 gluons show a clear EMC effect there is none in DSSZ. The EMC effect and strong antishadowing in the EPS09 gluons follow essentially from the PHENIX $\pi^0$ minimum-bias data \cite{Adler:2006wg} (see Fig.~\ref{fig:DY_DIS_pi0} above), which show these effects and which were given a fairly significant weight in the lack of other gluonic constraints in this region. As far as I can see, the explanation for the lack of these effects in the DSSZ gluons is that they are already fully included in the nFFs \cite{Sassot:2009sh} which DSSZ uses to compute the pion production cross sections. What seems troublesome to me here is that in the nFF analysis \cite{Sassot:2009sh} and in DSSZ one uses essentially the \textit{same} pion production data to fix the nuclear effects in the gluonic nFFs.  Since the nDS nPDFs, which do not show gluon antishadowing or EMC effects either, are utilized in the determination of these nFFs, essentially all the antishadowing and the beginning of the EMC effect in the measured pion  production are transferred into modifications of the gluonic nFFs. Then, by construction, these effects cannot anymore arise for the nPDFs in the DSSZ analysis, either. Clearly, this situation calls for a combined global analysis for the nFFs and nPDFs, which would exploit the different types of inclusive DIS, DY and pion production data together with the SIDIS data (see the Refs. in \cite{Sassot:2009sh}). In addition, one should also consider whether the very large nuclear effects obtained in the nFFs of Ref.~\cite{Sassot:2009sh} could be understood in terms of some feasible physical mechanism.

\begin{figure}[htbp]
\centering
\vspace{-0.3cm}
\includegraphics[width=8cm]{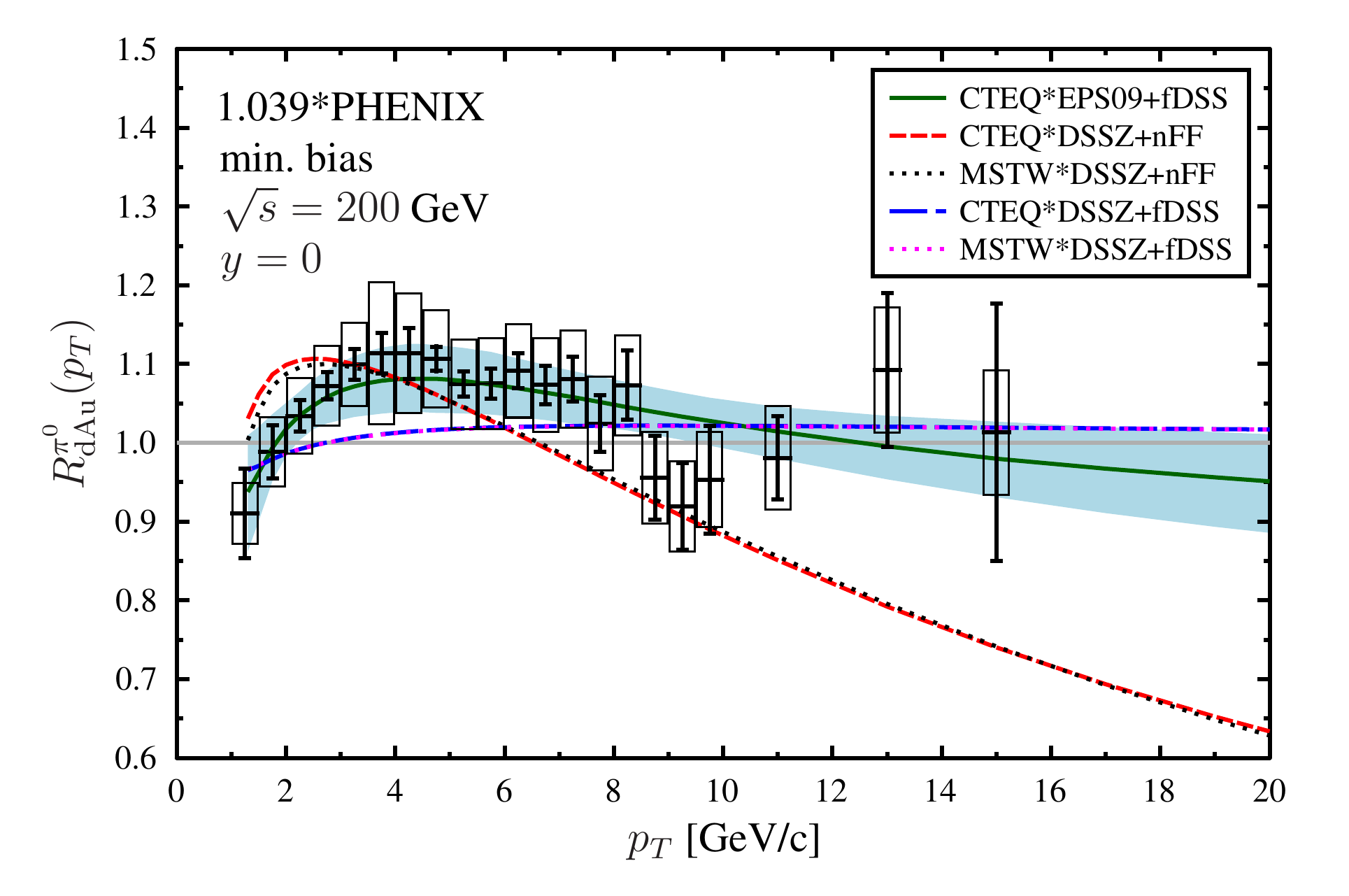}
\vspace{-0.5cm}
\caption{(Color on-line) The minimum-bias nuclear modification ratio $R_{\rm dAu}^{\pi^0}$ as a function of $p_T$,  see the text for details.
Prepared by I. Helenius.}
\label{fig:DSSZ_pi0}
\end{figure}
Figure~\ref{fig:DSSZ_pi0} illustrates the rather significant role the nFFs play in the DSSZ analysis. The PHENIX minimum-bias data for the nuclear modification factor $R_{\rm dAu}^{\pi^0}$ for $\pi^0$ production in d+Au collisions at RHIC, is scaled up by the factor 1.039 as in the EPS09 analysis (and also in Fig.~\ref{fig:DY_DIS_pi0}). The EPS09 result (green solid curve and error band) computed here with the CTEQ6M PDFs and fDSS vacuum FFs \cite{deFlorian:2007aj}, by construction fits the data. The corresponding result obtained with the DSSZ nPDFs and nFFS, using the CTEQ6M PDFs, is shown by the red dashed, and the result for DSSZ+nFF with MSTW by the black dotted curve (the p+p baseline in all these cases was computed with fDSS FFs). Unlike the nFFS, the free-proton PDFs are thus not a source of a large uncertainty here. The effect of changing the vacuum FFs into nFFs is, however, large: with DSSZ+vacuum FFs (blue dashed, purple dotted) essentially all the antishadowing and EMC effects in the computed $\pi^0$ production are now gone, indicating that all these effects are essentially included only in the (especially gluonic) nFFs in the DSSZ analysis.

\section{Impact-parameter dependent nPDFs: EPS09s and EKS98s}
\label{sec:spatial_nPDFs}

The spatial (impact parameter) dependence  arises naturally in the modeling of the origin for the nuclear shadowing, see e.g. \cite{Frankfurt:2011cs}, but  due to the lack of sufficient data constraints it has so far not been possible to embed such a dependence directly into the global analysis of nPDFs. However, for the computation of hard-process cross sections in different centrality classes (impact parameters \textbf{b}) of nuclear collisions, 
\begin{eqnarray}
\nonumber
\mathrm{d}N^{AB\rightarrow k + X}(\mathbf{b}) &=&  \sum\limits_{i,j,X'} \frac{1}{AB} \sum\limits_{N_A,N_B} \int \mathrm{d}^2\mathbf{s_1} \, T_A(\mathbf{s_1}) \, r_i^A(x_1,Q^2,\mathbf{s_1}) \, f_i^{N_A}(x_1,Q^2) \\ &\otimes& \int \mathrm{d}^2\mathbf{s_2} \, T_B(\mathbf{s_2}) \, r_j^B(x_2,Q^2,\mathbf{s_2}) \, f_{j}^{N_B}(x_2,Q^2) \otimes  \mathrm{d}\hat{\sigma}^{ij\rightarrow k + X'}
\delta^{(2)}(\mathbf{s_2} - \mathbf{s_1} - \mathbf{b}),
\label{eq:sigmaAB2}
\end{eqnarray}
one quite clearly needs to know also the spatial dependence of the nPDFs, i.e. how different the nPDFs are near the edge of the nucleus ($s\sim R_A$) from those at the center ($s\sim 0$). In the expression above, we have introduced the spatially dependent nuclear modification $r_i^A(x,Q^2,\mathbf{s})$ of the PDFs which is related to the average modification $R_i^A(x,Q^2)$ for each parton flavor $i$ by 
\begin{equation}
R_i^A(x,Q^2) \equiv \frac{1}{A}\int \mathrm{d}^2 \mathbf{s} \,T_A(\mathbf{s}) \,r_i^A(x,Q^2,\mathbf{s}),
\label{eq:normalization}
\end{equation}
where $T_A(\mathbf{s})$ is the standard nuclear thickness function. The indices $N_A(N_B)$ run through the nucleons in $A(B)$. 

As discussed by Helenius in these proceedings \cite{Helenius:2012wk}, further light on this issue was recently shed by the study in Ref.~\cite{Helenius:2012wd}, where the $A$ dependence of the EPS09 and EKS98 globally analysed nPDFs was turned into a spatial dependence through an ansatz involving powers of $T_A(\mathbf{s})$, 
\begin{equation}
r_i^A(x,Q^2,\mathbf{s}) = 1 + \sum\limits_{j=1}^{n} c^i_j(x,Q^2)\left[T_A(\mathbf{s})\right]^j.
\label{eq:ta_series}
\end{equation}
The main idea here is that the $A$-\textit{independent} coefficients $c^i_j(x,Q^2)$ are obtained for each $x$, each $Q^2$, each flavor $i$ of a particular nPDF set by performing a fit to reproduce the normalization in Eq.~(\ref{eq:normalization}) as a function of $A$. As shown in \cite{Helenius:2012wd} 
the (measured) $A$ systematics of the nPDFs can be reproduced nicely with $n=4$. For figures illustrating the resulting spatial dependence of the nPDFs, see Ref. \cite{Helenius:2012wd} and Fig. 2 in \cite{Helenius:2012wk}. The spatial fits were prepared for the EKS98 LO nPDFs, and for the EPS09 LO and NLO nPDFs as well as for all the EPS09 error sets. The outcome of this useful exercise, the spatially dependent nPDF sets named "EKS98s" and "EPS09s", is now publicly available\footnote{at https://www.jyu.fi/fysiikka/en/research/highenergy/urhic/nPDFs. Instructions of how to straightforwardly implement these nPDFs in the existing codes for the hard process cross sections are given in Ref.~\cite{Helenius:2012wd}.}. With these, one may now for the first time compute nuclear hard-process cross-sections and estimate their nPDF-originating uncertainties in different centrality classes (or impact parameters) consistently with a globally analysed nPDFs. 

As shown in Ref.~\cite{Helenius:2012wd} and discussed by Helenius in these proceedings \cite{Helenius:2012wk} (see Fig. 3 there), with the EPS09s nPDFs one can, within the experimental and theoretical uncertainties, nicely reproduce the PHENIX data \cite{Adler:2006wg} on the centrality dependence of the nuclear modification factor $R_{\rm dAu}^{\pi^0}$. Predictions for the centrality dependence of the corresponding modification in the forthcoming p+Pb collisions at $\sqrt s_{NN}= 5.0$ TeV at the LHC have also been computed with EPS09s, see Ref.~\cite{Helenius:2012wd}, and Fig.~4 in \cite{Helenius:2012wk}.

\section{Conclusions}
\label{sec:conclusions}
Based on the recent global analyses of nPDFs, I conclude the following: 
\textit{(i)} The very good quality fits obtained suggest that in the $x,Q^2$-region probed by the data, $x \gtrsim 0.001$ and $Q^2 \gtrsim 1$~GeV$^2$, factorization works and the nPDFs seem universal.
\textit{(ii)}
The NLO analyses with error sets (EPS09, DSSZ) have brought the nPDF global fits to the similar (NLO) sophistication level as the free-proton PDF analyses.
\textit{(iii)}
There are still large uncertainties in the gluon sector: The role of the nuclear FFs in understanding $R_{\rm dAu}^{\pi^0}$ should be clarified by performing a simultaneous global fit of nPDFs and nFFs.
\textit{(iv)}
The free-proton PDF uncertainties are so far only partially (or not at all) accounted for in the nPDF analyses. Ultimately, one should develop a combined global analysis of PDFs and nPDFs.
\textit{(v)}
Further hard-process data from RHIC d+Au and LHC p+Pb collisions, such as direct photons, high-$p_T$ pions, heavy quark+photon production and 
$Z/W$ asymmetries, will help in constraining the nPDFs further.
\textit{(vi)}
To resolve the gluon uncertainties at small-$x$\&high-$Q^2$, DIS data from future $e$+$A$ colliders (EIC, LHeC) would be needed.
\textit{(vii)}
The impact parameter dependence of globally analysed nPDFs has now been determined (in a specific framework) for the first time consistently with their $A$-systematics: with EPS09s (NLO\&LO + error sets) and EKS98s (LO) it is now possible to compute nuclear hard-process cross-sections in different centrality classes consistently with EPS09 and EKS98. A future task is to implement this type of framework directly into the global fits.
\vspace{0.3cm}

\noindent
{\bf Acknowledgments.} I thank Ilkka Helenius for preparing the Figs.~\ref{fig:DSSZvsEPSinx} and \ref{fig:DSSZ_pi0} above, T. Lappi and I. Helenius for comments regarding the talk and this manuscript, and DSSZ for providing us with the DSSZ nPDFs.
The financial support from the Academy of Finland, project 133005, and Magnus Ehrnrooth Foundation is gratefully acknowledged.





\bibliographystyle{elsarticle-num}
\bibliography{HP2012proc_eskola}







\end{document}